\documentclass[aps,a4paper,10pt,twocolumn,nofootinbib]{revtex4} 
\usepackage{amsfonts}
\usepackage{amssymb}
\usepackage{mathrsfs}
\usepackage[mathscr]{euscript}
\usepackage[dvips]{graphicx}
\usepackage{fancyhdr}

\begin{document}

\title{Phase coherence and extreme self phase modulation}
\author{P. Kinsler}
\email{Dr.Paul.Kinsler@physics.org}
\affiliation{
  Blackett Laboratory, Imperial College London,
  Prince Consort Road,
  London SW7 2AZ,
  United Kingdom.}

\begin{abstract}

I study how pulse to pulse phase coherence in a pulse train 
 can survive super-broadening by extreme self phase modulation (SPM).
Such pulse trains have been used in phase self-stabilizing schemes
 as an alternative to using a feedback process.
However, 
 such super-broadened pulses have undergone considerable distortion, 
 and it is far from obvious that they necessarily 
 retain any useful phase information.
I propose measures of phase coherence applicable to such pulse trains, 
 and use them to analyze numerical simulations comparable to 
 self-stabilization experiments.

\end{abstract}



\newcommand{\sech}{{\textrm{ sech}}}

\lhead{\includegraphics[height=5mm,angle=0]{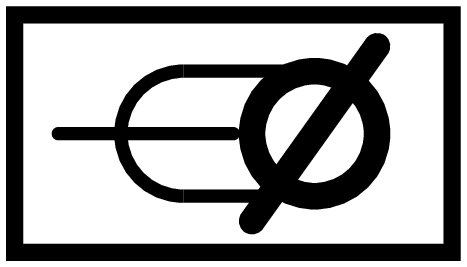}~~PHSPM}
\chead{~}
\rhead{
\href{mailto:Dr.Paul.Kinsler@physics.org}{Dr.Paul.Kinsler@physics.org}
}
\lfoot{}
\rfoot{Kinsler-2007-phspm}

\date{\today}
\maketitle
\thispagestyle{fancy}


\chead{Phase coherence in XSPM}

%
\section{Introduction}

In some  recent experiments 
 (e.g. \cite{Fang-K-2004ol,Baltuska-FK-2002prl}),
 carrier-envelope phase (CEP) stabilized idler pulses 
 were generated in an optical parametric amplifier (OPA) 
 by combining a pump pulse with
 its super-broadened replica as the signal pulse.
The phase relationships in the optical parametric interaction
 ensure that the phase of the idler pulse 
 is independent of that of the pump.  
This principle has been tested experimentally 
 using a train of pump pulses with arbitrary CEP \cite{Fang-K-2004ol}, 
 and a train of idler pulses with stabilized phase has been achieved.
The distinguishing feature of this technique is that it acts
 on each pulse in isolation, 
 and does not rely on feedback.
The idler phase is stabilized to an unknown value
 determined by the details of the spectral broadening process.

\begin{figure}[ht]
\begin{center}
 \includegraphics[width=0.960\columnwidth,angle=-0]{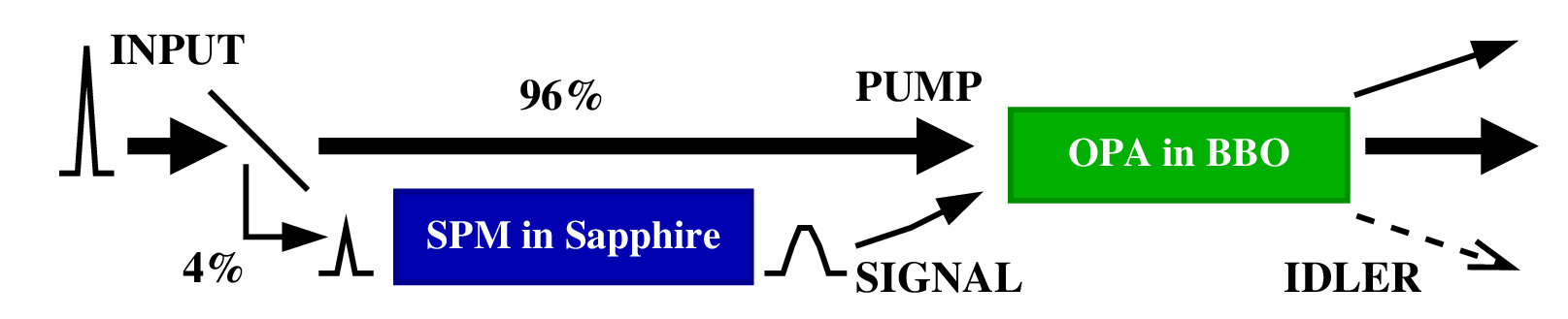}
\end{center}
\caption{
Diagram of the Fang et al. experiment \cite{Fang-K-2004ol}.  
Pulses of unknown CEP from an input train have a small fraction 
 split off at a beam splitter and then broadened in sapphire.
The rest of the input pulse (i.e. the pump pulse) is then combined 
 with the broadened portion, 
 in a non-collinear OPA,
 generating a phase stabilized idler pulse.
}
\label{fig-fangexpt}
\end{figure}

Fang et al. \cite{Fang-K-2004ol} broadened their signal pulses
 using extreme self-phase modulation (SPM) in sapphire, 
 then selected suitable spectral components
 from the super-broadened spectrum to form the signal pulse
 for their OPA stage (see fig. \ref{fig-fangexpt}).
The success of the experiment raises interesting questions
 about how the severely distorted temporal profile of the signal pulse
 can contain sufficient sensible phase information
 to achieve the desired result, 
 especially given that small differences between input pulses can 
 be turned into large differences by the strong nonlinearity.

In this paper I investigate how (inter-pulse) phase coherence 
 can survive in a train of pulses broadened by a 
 third-order nonlinearity.
In most pulse propagation models based on 
 a complex envelope $A(t)$, 
 the nonlinear interaction term governing 
 self-phase modulation (SPM) 
 appears as $\chi^{(3)} |A|^2 A$.  
This is linear in phase (of $A$), 
 and is usually expected to be well-behaved, 
 although intensity fluctuations in the pulse train 
 lead to CEP fluctuations in the output
 \cite{Baltuska-UGKYUHF-2003jstqe,Li-MWMNTC-2007ol}.
However, 
 the true form of the nonlinear interaction is 
 $\chi^{(3)} E^3$.
In the envelope picture, 
 this adds an extra (non resonant and CEP sensitive) term to the 
 nonlinearity, 
 i.e. $\chi^{(3)} A^3$.
This initially leads to third harmonic generation (THG) 
 which can disrupt the phase stabilization process; 
 higher (odd) harmonics can also be generated.
Even more importantly, 
 the THG is generated from the whole broadened spectrum, 
 not just the initial central peak.
Thus we cannot always assume that all of the THG contribution 
 will remain safely out of range of our SPM broadened part:
 we might well expect that 
 fluctuations will be strongly mixed into the nonlinear process, 
 destroying any pulse-to-pulse repeatability.
Since the true nonlinearity includes this wide range
 of multi-wave interactions, 
 we need to be aware that 
 all have the potential to obscure 
 the phase coherence of the train of broadened pulses.


I define criteria characterizing the effect 
 of fluctuations in intensity, 
 pulse width, 
 and CEP on the phase of the broadened pulse.
Applying these to simulations comparable to the experiment
 of Fang et al.,
 I show the importance of minimising fluctuations in the 
 input pulse train:
 even though SPM is insensitive to CEP variation,
 intensity and width fluctuations can play a significant role.
The wide bandwidth of these pulses, 
 along with the need to retain the full $\chi^{(3)}$ interaction, 
 places significant demands 
 on the required numerical resolution.
Therefore simulations of pulse propagation were done using 
 the PSSD method\cite{Tyrrell-KN-2005jmo}, 
 which evolve the carrier oscillations directly.

Because the strong nonlinearity means that the results will vary
 significantly from one set of parameters to another,
 the data presented in this paper serves largely
 to indicate the various issues involved in extreme SPM
 broadening, 
 rather than being a set of specific predictions.
Nevertheless, 
 the simulation results indicate that the phase coherence
 in the broadened pulse train
 is maintained only up to a point, 
 and that sensitivity to pulse intensity or width fluctuations
 is much stronger than for CEP variation.
Further, 
 SPM can only broaden pulses by so much 
 before multi-wave cross talk and/or THG
  destroy the pulse-to-pulse phase coherence.

\noindent
In section \ref{S-phaseproperties} the basic phase properties
 relevant for this analysis are described, 
 and in section \ref{S-phasecoherence} 
 the measures of phase coherence are introduced. 
Section \ref{S-simulations} describes the simulation methods used, 
 and then in section \ref{S-variation} the effects of fluctuations
 in the input pulse train
 (intensity, pulse width, and CEP) are described.
Finally, 
 section \ref{S-discussion} comprises a brief discussion, 
 and section \ref{S-conclusions} contains my conclusions.

%
\section{Phase properties}
\label{S-phaseproperties}

Here I focus on the spectral phase $\phi_s(\omega)$, 
 since a broadened pulse will suffer chirp and other distortion,
 removing any straightforward way to characterize the phase in the time domain.
This spectral phase will
 depend on the phase of the input pulse $\phi_p(\omega)$, 
 on its other properties (intensity, width, wavelength, etc), 
 and on the propagation medium (e.g.  sapphire).
Considering just intensity $I$ and input CEP $\phi_p$ variation,
 we have that the (broadened) signal pulse has a phase
~
\begin{eqnarray}
  \phi_s(I,\phi; \omega) 
&=&
  \phi_p(\omega) + \Phi(I,\phi; \omega) 
\\
&=& 
  \phi_p(\omega) 
 +
  \Phi(I_0,\phi_0; \omega) 
\nonumber
\\
&& ~~~~
 +
  \Delta_I(I_0,\phi_0) (I-I_0)  
 +
  \delta_I(\omega)
\nonumber
\\
&& ~~~~
 +
  \Delta_\phi(I_0,\phi_0) (\phi-\phi_0)  
 +
  \delta_\phi(\omega)
,
\label{eqn-fluctphispm-general}
\end{eqnarray}
where $I_0$ is the average intensity of pulses in the train,
 and $\phi_0$ the average CEP; 
 further terms including the effect of pulse width ($\tau$) variation 
 can easily be added.
The function $\Phi$ encodes the full effect 
 of the complicated nonlinear interactions on the phase.
Here the linear responses of the phase about 
 $I_0$ and $\phi_0$ are given by $\Delta_I$ and $\Delta_\phi$, 
 while the remaining nonlinear response is contained 
 in $\delta_I, \delta_\phi$.

The broadened signal pulse is then mixed with the 
 original pulse in an OPA.
The difference-frequency idler pulse that results has a spectral 
 phase which is the difference of the pump phase $\phi_p(\omega)$
 and the signal phase $\phi_s(\omega)$, 
~
\begin{eqnarray}
  \phi_i(I,\phi; \omega)
&=&
  \phi_p(\omega)-\phi_s(I,\phi; \omega)
\\
&=& 
 - 
  \Phi(I_0,\phi_0; \omega) 
\nonumber
\\
&& ~~~~
 -
  \Delta_I(I_0,\phi_0) (I-I_0)  
 -
  \delta_I(I,\phi; \omega)
\nonumber
\\
&& ~~~~
 -
  \Delta_\phi(I_0,\phi_0) (\phi-\phi_0)  
 -
  \delta_\phi(I,\phi; \omega)
.
\label{eqn-fluctphiopa-intensity}
\end{eqnarray}
We can see that each idler pulse output from the OPA
 will have a phase $\phi_i$ stabilized to a value set by 
 the nonlinear propagation through the sapphire --
 the unknown CEP $\phi_p(\omega)$ of the input pump pulse has 
 has canceled out.
However, 
 this phase $\phi_i$ will vary from pulse to pulse 
 depending on how the fluctuations 
 in the input pulse train
 alter each instance of the linear and nonlinear evolution.
Although 
 I consider only the (three-wave) difference-frequency generating term, 
 note that the full $\chi^{(2)}$  OPA interaction 
 contains potentially phase sensitive terms
 \cite{Kinsler-NT-200X-phnlo}.

%
\section{Phase coherence}
\label{S-phasecoherence}

In order to determine whether there is sufficient
 useful pulse-to-pulse phase coherence after the 
 SPM broadening, 
 I analyze the last four terms in eqn.(\ref{eqn-fluctphispm-general}) 
 to define the necessary measures of phase coherence.

I first consider the {\em linear response} of the 
 output pulse to a variation in the input pulse;
 e.g. how the spectral phase of the pulse varies due to
 small changes in the intensity. 
Knowing the linear response enables us to set tolerances on the 
 input pulse train to guarantee a certain level of 
 stability between pulses in the output train.
We also 
 need to know how reliable this estimated linear response is: 
 since we are considering a complex nonlinear interaction, 
 we cannot simply the response
 to input variations will be linear.
What we hope is that the dominant response will be linear, 
 and that the nonlinear corrections will remain small.

The quantities defined below in 
 eqns. (\ref{eqn-coherence-deltaI}) and (\ref{eqn-coherence-VI})
 are expressed for simplicity only w.r.t. intensity
 fluctuations; 
 but by swapping the intensity arguments $I_1, I_2$ 
 for (e.g.) CEP's  $\phi_{p1}, \phi_{p2}$,
 we can equally well define a $\Delta_\phi, \sigma^2_\phi$, 
 expressing the response to input CEP variation; 
 just as by swapping the intensity $I_1, I_2$ 
 for pulse widths $\tau_1, \tau_2$
 we could define $\Delta_\tau, \sigma^2_\tau$.

Since our simulations provide data (such as spectral phases) 
 not easily accessible in experiments, 
 we can use this to our advantage in determining the likely coherence 
 properties of the broadened pulse train, 
 and even suggest changes to improve performance.

%
\subsection{Linear response}
\label{Ss-coherence-avgvarphi}

The linear response of the phase to intensity variation
 can be estimated by taking an ensemble average over 
 a large number of simulations.
Each simulation starts
 with a pulse selected at random from the 
 distribution of all possible pulses in 
 the input pulse train.
I define the linear response as
 an ensemble average over pairs of broadened signal pulses:
~
\begin{eqnarray}
  \Delta_I(\omega)
&=&
  \left<
    \frac{ \phi_s(I_1;\omega) - \phi_s(I_2;\omega) }
         { \left| I_1 - I_2 \right| }
  \right>
.
\label{eqn-coherence-deltaI}
\end{eqnarray}
More simply, 
 we might calculate $\Delta_I$ by simply 
 taking an average of phase differences over a 
 suitable range of pulse variation.
If this $\Delta_I$ is small, 
 phase is little affected by input variation --
 at least to a linear approximation.
For the case of intensity variation, 
 this linear response has been measured experimentally, 
 as reported by Li et al. \cite{Li-WMNTC-2007ol}.

%
\subsection{Nonlinear response}
\label{Ss-coherence-varphi}

I estimate the contribution of the nonlinear part of the 
 phase response using
~
\begin{eqnarray}
  \sigma^2_I (\omega)
&=&
  \left<
    \delta^2_I(\omega)
  \right>
~
=
  \left<
    \frac{ \left[ \phi(I_1;\omega) - \phi(I_2;\omega) \right]^2 }
         { \left( I_1 - I_2 \right)^2 }
   -
    \Delta_I(\omega)^2
  \right>
.
~~~~
\label{eqn-coherence-VI}
\end{eqnarray}
Although $\sigma^2_I(\omega)$ is calculated in the same manner as a variance,
 it does not mean that $\phi_s$ varies randomly as
 the intensity (or some other parameter) shifts.
This $\sigma^2_I(\omega)$ is simply a measure of how nonlinear 
 the response of  $\phi_s$ is to pulse variation.

If this $\sigma^2_I(\omega)$ is small ($\sigma_I \ll \Delta_I$), 
 then the phase changes due to pulse fluctuations is predominately linear, 
 and $\Delta_I$ is a useful quantity.
Note that the size of $\sigma^2_I(\omega)$ is strongly dependent 
 on the range of variation in $I$; 
 if that range is altered then $\sigma^2_I(\omega)$ should be recalculated.

%
\subsection{Other measures of coherence}
\label{Ss-coherence-dudleyc}

The measures I introduce above 
 could not be easily measured in an experiment:
 although the spectral phase of pulses can be measured 
 (e.g. \cite{Kobayashi-SF-2001jqe}), 
 doing this for each individual pulse in a train
 of (dissimilar) broadened pulses
 would be a rather challenging task.
Here they are intended primarily as a theoretical construct useful for 
 analyzing simulation results.
In contrast, 
 Dudley and Coen \cite{Dudley-C-2002ol} have proposed 
 a measure of ``shot to shot'' coherence $g_{12}^{(1)}$, 
 which can be calculated by taking an ensemble average 
 of pairs of results taken from a set of simulations.
The measure is
~
\begin{eqnarray}
  \left| 
    g_{12}^{(1)} (\lambda,t_1-t_2)
  \right|
&=&
  \left|
    \frac{ \left< E_1^*(\lambda,t_1) E_2(\lambda,t_2) \right>}
         { \sqrt{ 
              \left< \left|E_1(\lambda,t_1)\right|^2 \right>
              \left< \left|E_2(\lambda,t_2)\right|^2 \right> 
         } }
  \right|
.
\end{eqnarray}
Although this is easy to calculate from a simulation,
 it includes contributions from intensity variation
 in the broadened pulses, 
 a complication which we need to avoid in this work.

%
\section{Simulations}
\label{S-simulations}

The simulations were done using
 the PSSD method\cite{Tyrrell-KN-2005jmo,Fornberg-PSmethods}, 
 which offers significant advantages over the 
 traditional FDTD 
 and Pseudospectral Time-Domain (PSTD) techniques 
 (see e.g. \cite{Gilles-HV-2000jcp})
 for modeling the propagation and interaction of few-cycle pulses.
Run times are generally faster, 
 and the PSSD method also offers far greater flexibility 
 in the handling of dispersion.
Whereas FDTD \& PSTD propagate fields $E(z), H(z)$ 
 forward in  time, 
 PSSD propagates fields $E(t), H(t)$ forward in space.
Under PSSD, 
 the entire time-history (and therefore frequency content) 
 of the pulse is known at any point in space, 
 so arbitrary dispersion incurs no extra computational penalty.
In contrast, the FDTD or PSTD approaches must use convolutions 
 or time-response models for dispersion.

However, 
 since $z$-propagated simulations do not handle either reflections
 or backward waves easily, 
 we need to ensure we remain in the uni-directional propagation limit, 
 where for an $n$-th order perturbative nonlinearity, 
 $n \chi^{(n)} E^{n-1} / n_0^2 \ll 1$ (see e.g. \cite{Kinsler-2007josab}).
Note also that it is also possible to do these simulations using 
 explicitly directional fields
 \cite{Kinsler-RN-2005pra,Kolesik-M-2004pre,Kinsler-2007arXiv-envel} 
 or even wideband envelopes \cite{Genty-KKD-2007ox,Kinsler-2007arXiv-envel}.

I apply the PSSD algorithm to the
 source-free Maxwell's equations in non-magnetic media, 
 with an instantaneous $\chi^{(3)}$ nonlinearity,
 so that 
 the equations for $E$ and $H$ in the 1D (plane wave) limit are
~
\begin{eqnarray}
  \frac{dH_y(t;z)}{dz} 
&=&
 -
  \frac{d}{dt}
  \left[
    \epsilon_0 \epsilon_r (t) * E_x(t;z) + \epsilon_0 \chi^{(3)}  E_x(t;z)^3
  \right]
,
\label{eqn-pssd-dH}
\\
  \frac{dE_x(t;z)}{dz} 
&=&
 -
  \frac{d}{dt}
  \left[
    \mu_0 H_y(t;z)
  \right]
,
\label{eqn-pssd-dE}
\end{eqnarray}
 where the $*$ denotes the convolution necessary to allow for 
 the dispersion of the medium.

Typical array sizes used were $N=2^{15}$, 
 with time resolution of 0.1fs (hence a time window of $T=3.2768$ps).
Spatial propagation steps were chosen to be $dz = 0.4 c T/N \approx 12$nm
 in order to ensure numerical stability.
The pulse profile used as an initial condition was
~
\begin{eqnarray}
  E(t) 
&=&
  E_0 
  \sin(\omega_p t + \phi) 
  \sech(t/\tau)
.
\label{eqn-initialfield}
\end{eqnarray}
Pulses were propagated through 2mm of sapphire modeled 
 using the dispersion parameters from 
 DeFranzo \& Pazol \cite{DeFranzo-P-1993ol},
 and nonlinearity data from Major et al. \cite{Major-YNAS-2004ol}.
The pulse wavelength was $\lambda_p = 786$nm, 
 which corresponds to $\omega_p \simeq 2.4\times 10^{15}$rad/s.
The reference peak intensity $I_{ref}$ for the part of the 
 pulse being SPM broadened was chosen to be compatible with the
 pulse energy of 4\% of 1.5mJ (length 130fs) reported 
 by Fang et al. \cite{Fang-K-2004ol}, 
 being
 $I_{ref} = 0.33 \times 10^{14}$W/cm$^2$.
In \cite{Fang-K-2004ol}, 
 their OPA stage selected a 
 a signal wavelength of $\lambda_s \sim 1400$nm 
 by angle tuning, 
 generating an idler  $\lambda_i \sim 1600$nm.
Here, 
 however, 
 we simulate only the important SPM broadening stage.

It is worth noting that in these simulations,
 a transform-limited 130fs pulse is seen to be
 too narrowband to undergo suffient broadening before multi-wave
 cross-talk destroyed the phase coherence.
Consequently, 
 we shortened our default pulse length to 30fs, 
 as the simpler alternative to adding a strong chirp.

%
%
%
%
%

The simulations did not incorporate transverse effects
 in order to keep computation times down.
Each 1D simulation through 2mm of sapphire takes about one hour
 on a 2.4GHz PC, 
 so individual simulations involving transverse effects
 would in principle be tractable over days or weeks.
However,
 the results here depend on multiple simulation sets: 
 a small number of 3D simulations  
 could not have addressed the concerns of this paper.

%
\section{Fluctuations in the pulse train}
\label{S-variation}

Some 
 previous work \cite{Tyrrell-2006phd,Kinsler-RTN-2006cleo}
 showed how pulses broadened in experiments 
 like those of Fang et al.\cite{Fang-K-2004ol}
 might preserve useful phase information.
However, 
 these simulations used artificial nonlinear strengths and pulse widths
 to ensure the presence of a strong SPM lobe directly at the desired 
 signal frequency in order to guarantee good performance.
They also included the OPA stage,
 demonstrating that it could preserve the 
 phase information supplied by the SPM lobe.
These indicated a phase stability at the output 
 of the OPA stage similar to that seen by 
 Fang et al., 
 even though the parameters used were not strictly 
 comparable.

I now present simulation results that match the pulse intensity and 
 material dispersion more closely.
Fang et al.
 were unlikely to have been operating in the 
 regime where a strong SPM lobe sat at their chosen signal frequency,
 since that was far into the wings of the broadened spectra.
Unfortunately, 
 the match is not perfect since 
 the experimental parameters quoted by 
 Fang et al. 
 are insufficient to fully characterize a simulation.
Fang et al. used a near-degenerate OPA setup, 
 so were looking to broaden their input pulses from 2.4rad/fs 
 all the way down to 1.2rad/fs; 
 thus we are most interested in the results in the nearby frequency range
 $\omega \sim 1.3$ rad/fs

It is therefore difficult to be sure 
 whether a phase-stable train of broadened pulses will arise, 
 since
 we need the wing of the broadened spectra to be 
 stronger than the underlying background of complicated 
 phase-sensitive nonlinear processes.
When doing the large set of simulations representing the 
 variability of the input pulse train, 
 three different pulse properties were altered in turn:
 intensity, 
 pulse width, 
 and CEP.
We chose variations of the order of a few percent for 
 intensity and pulse width:
 specifically a range of 5\% in steps of 0.5\%.
For CEP, 
 we obtained results for a $\pi/2$ range in CEP in 
 steps of $\pi/16$, 
 which provides results applicable for the full $2\pi$.

\begin{figure}
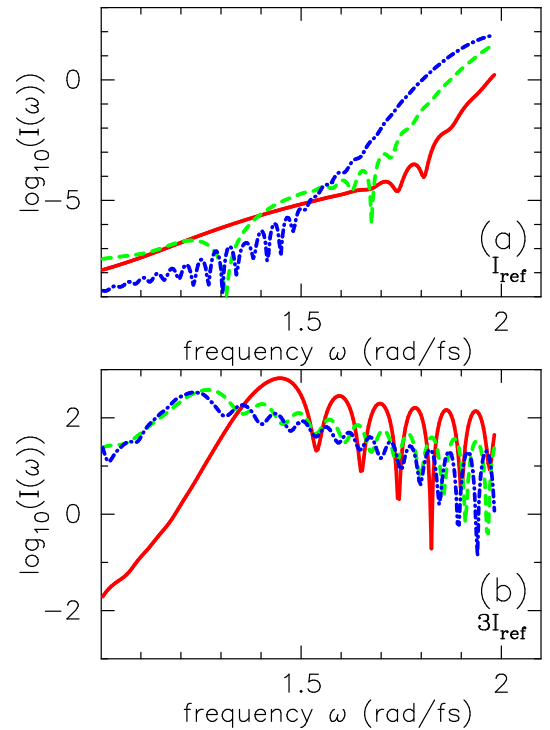
 
\begin{center}
 \includegraphics[angle=-90,width=0.80\columnwidth]{fig02a-spectra-a1}\\
 \includegraphics[angle=-90,width=0.80\columnwidth]{fig02b-spectra-a3}
\end{center}
\caption{
Spectral intensities for
(a) $I_0=I_{ref}=0.33\times 10^{14}$W/cm$^2$, 
(b) $I_0=3I_{ref}$; 
at 
(solid line, red, medium) 0.85mm, 
 (dashed line, green, light) 1.43mm,
 (dot-dashed line, blue, dark) 2.00mm.
Peak intensity values at $z=0$mm are about $10^3$.
}
\label{fig-pcintensity-Iw}
\end{figure}

The spectral intensities 
 at our chosen input intensities and propagation distances 
 are shown on fig. \ref{fig-pcintensity-Iw}.
We see that the lowest input intensity pulse has only
 minimal power present in the range of interest,
 even for the full 2mm propagation, 
 whereas the highest intensity pulse broadens rapidly.
Further, 
 we see (in the high intensity case) the 
 usual series of SPM lobes pushing down to lower 
 frequencies as the pulse propagates.
We need to bear these spectra in mind when analyzing 
 the following results, 
 since a well behaved phase response will be of little practical use
 if those frequency components are undetectable.
This caveat is particularly relevant in the lowest pulse intensity
 cases.

Note that in all the following figures, 
 the central frequency of the input pulse
 is off the right hand side of the frame, 
 since $\omega_p \simeq 2.4$rad/fs.

%
\subsection{Intensity fluctuations}
\label{ss-variation-intensity}

Pure SPM causes a phase shift proportional to intensity,
 but
 here I am instead interested in the unavoidable 
 and potentially significant complications
 arising because the nonlinearity
 also includes THG effects.

\begin{figure}
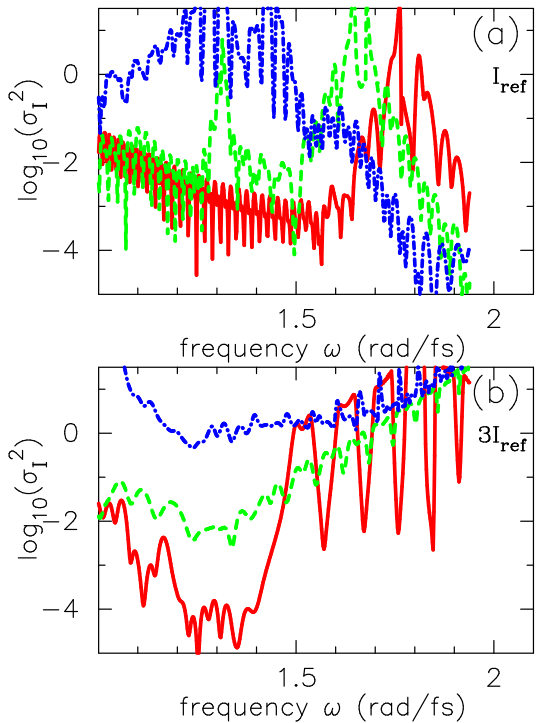
 
\begin{center}
 \includegraphics[angle=-90,width=0.80\columnwidth]{fig03a-vardelphi-amplitude-a1.ps}\\
 \includegraphics[angle=-90,width=0.80\columnwidth]{fig03b-vardelphi-amplitude-a3.ps}
\end{center}
\caption{
The effect of intensity fluctuations on the phase coherence measure
for (a) $I_0=I_{ref}=0.33\times 10^{14}$W/cm$^2$,
(c) $I_0=3I_{ref}$;
(solid line, red, medium) 0.85mm, 
 (dashed line, green, light) 1.43mm,
 (dot-dashed line, blue, dark) 2.00mm.
A value of $\log(\sigma^2_I)=-2$ corresponds to a 
     nonlinear phase adjustment $\delta_I \sim \pm 0.1$ rad / \%.
  Width $\tau=30$fs;  
  $\lambda_p=786$nm; $\rightarrow \omega_p \simeq 2.4 $rad/fs.
}
\label{fig-pcintensity-3}
\end{figure}

Fig. \ref{fig-pcintensity-3} 
 plots the logarithm of the phase variance $\sigma^2_I$
 cause by intensity fluctuations, 
 for three propagation distances at two different intensities.
Linear phase response only holds to a level of 0.1rad/\% input intensity 
 fluctuations where the
 log-variance is less than $-2$.
In these results, 
 it is important to note that if the intensity fluctuates, 
 the central frequencies of the SPM lobes also shift, 
 and this can cause large changes in the phase variance $\sigma^2_I$.

First consider the results for the reference 
 intensity $I_0=I_{ref}$, fig. \ref{fig-pcintensity-3}(a).
By comparing the $\log_{10} \sigma^2_I$ curves at different distances, 
 we see that the trend at higher frequencies is for the 
 phase variances to decrease, 
 a result of the SPM broadened spectral peak pushing outwards.
In contrast, 
 for lower frequencies the opposite trend occurs, 
 because the gradually increasing background of phase-sensitive 
 processes still tends to dominate the spectral wings of the pulse.
Indeed, 
 for the 2mm propagation distance, 
 the variances have increased to the point where 
 any hope of a linear phase response has been lost.
Note the spike in $\sigma^2_I$ for 1.3rad/fs at 1.43mm, 
 this corresponds to the dip in intensity seen on 
 fig. \ref{fig-pcintensity-Iw}(a).

Similar trends are shown
 for the higher intensity 
 $I_0=3I_{ref}$, fig. \ref{fig-pcintensity-3}(c), 
 although the centre of the spectrum rapidly loses phase coherence, 
 largely due to the strong SPM-induced spectral modulation.
Note also the significant modulation of the variances caused by the 
 SPM lobes in the intensity spectrum, 
 which are clearly visible early in the propagation at 0.85mm.

\begin{figure} 
\begin{center}
 \includegraphics[angle=-90,width=0.80\columnwidth]{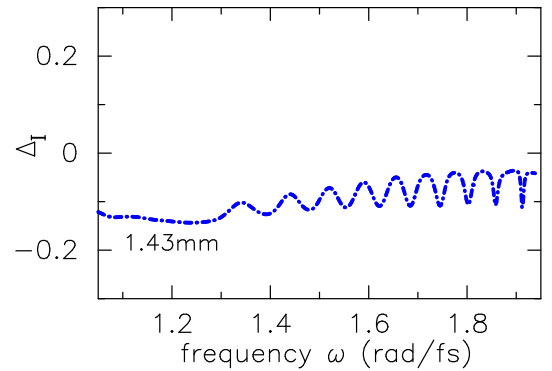}
\end{center}
\caption{
The linear response of the phase to intensity fluctuations at 
1.43mm and 
 (dot-dashed blue line, dark) $I_0=3I_{ref}$.
  Width $\tau=30$fs;  
  $\lambda_p=786$nm; $\rightarrow \omega_p \simeq 2.4$rad/fs.
}
\label{fig-pcintensity-avg-3}
\end{figure}

For both intensities, 
 there is a window of low phase variance at 1.43mm, 
 although only for the $I_0=3I_{ref}$ case is there also appreciable 
 spectral intensity in this region.
Therefore on 
 fig. \ref{fig-pcintensity-avg-3} we can plot $\Delta_I$ 
 this ($3I_{ref}$) case.
For an $\omega_s \simeq 1.3$rad/fs, 
 we see that $\Delta_I \simeq -0.1$ rad/\%, 
 hence phase stability is maintained to less than $0.1$rad
 if the intensity variation between the train of input pulses 
 is less than 1\%.
However, 
 in this particular case, 
 the linear phase shift (for a 1\% variation) is of the same order
 as the nonlinear contribution.

%
\subsection{Width fluctuations}
\label{ss-variation-width}

\begin{figure}
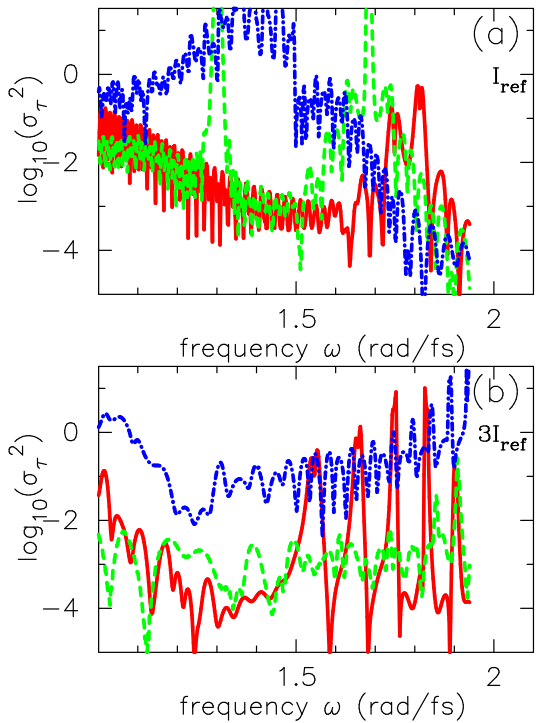
 
\begin{center}
 \includegraphics[angle=-90,width=0.80\columnwidth]{fig05a-vardelphi-width-a1.ps}\\
 \includegraphics[angle=-90,width=0.80\columnwidth]{fig05b-vardelphi-width-a3.ps}
\end{center}
\caption{
The effect of pulse width fluctuations on the phase coherence measure.
for (a) $I_0=I_{ref}=0.33\times 10^{14}$W/cm$^2$,
(c) $I_0=3I_{ref}$;
(solid line, red, medium) 0.85mm, 
 (dashed line, green, light) 1.43mm,
 (dot-dashed line, blue, dark) 2.00mm.
A value of $\log(\sigma^2_\tau))=-2$ corresponds to a 
     nonlinear phase adjustment $\delta_\tau \sim \pm 0.1$ rad / \%.
  Average width $\tau=30$fs;  
  $\lambda_p=786$nm; $\rightarrow \omega_p \simeq 2.4$rad/fs.
}
\label{fig-pcwidth-3}
\end{figure}

Fig. \ref{fig-pcwidth-3}
 shows the effect of pulse width fluctuation on 
 the output phase response.
Since the width fluctuations correspond to bandwidth fluctuations, 
 they also cause different amounts of SPM-induced broadening, 
 and shifting SPM lobe positions; 
 hence we see similar trends as for intensity fluctuations.

Again we see the spike at $I_0=I_{ref}$, 1.3rad/fs and 1.43mm, 
 caused by the dip in the intensity spectra.
Also, 
 the variances tend to increase with propagation distance.
Interestingly, 
 the $I_0=3I_{ref}$ variances are smaller than those for the 
 lower intensity $I_0=I_{ref}$ case, 
 giving an example of how the balance 
 between phase-sensitive nonlinear effects
 and coherent (SPM) spectral broadening can change.
More generally, 
 however, 
 variations of $\sim 1$\%) in width or intensity have a 
 comparable effect, 
 although there are many differences of detail.

%
\subsection{CEP fluctuations}
\label{ss-variation-phase}

The final type of pulse variation we consider is CEP fluctuations.
This case differs markedly from those for variations in intensity or width, 
 since those were dominated by the concomitant alterations to the SPM.
In contrast, 
 since SPM is CEP insensitive, 
 CEP fluctuations can only act (at least initially) 
 through the THG-like contribution 
 to the nonlinearity.

\begin{figure}
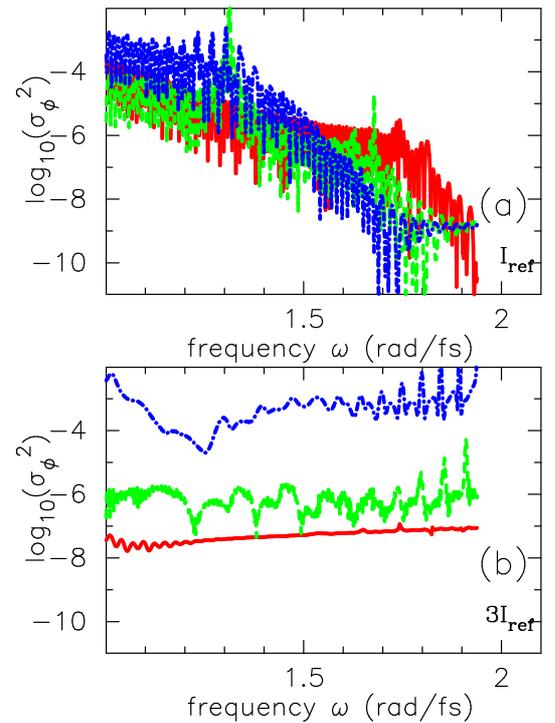
 
\begin{center}
 \includegraphics[angle=-90,width=0.80\columnwidth]{fig06a-vardelphi-phase-a1.ps}\\
 \includegraphics[angle=-90,width=0.80\columnwidth]{fig06b-vardelphi-phase-a3.ps}
\end{center}
\caption{
The effect of input phase fluctuations on the phase coherence measure.
for (a) $I_0=I_{ref}=0.33\times 10^{14}$W/cm$^2$,
(c) $I_0=3I_{ref}$;
(solid line, red, medium) 0.85mm, 
 (dashed line, green, light) 1.43mm,
 (dot-dashed line, blue, dark) 2.00mm.
A value of $\log(\sigma^2_\phi))=-2$ corresponds to a 
     nonlinear phase adjustment $\delta_\phi \sim \pm 0.1$ rad / \%.
  Width $\tau=30$fs;  
  $\lambda_p=786$nm; $\rightarrow \omega_p \simeq 2.4$rad/fs.
}
\label{fig-pcphase-3}
\end{figure}

The results in fig. \ref{fig-pcphase-3} show that 
 even for the high intensity ($I_0=3I_{ref}$) case, 
 the response of the output spectral phase to the 
 full range of CEP variation in the input pulse 
 is remarkably linear.
This emphasizes that the CEP stabilization scheme can 
 work as expected, 
 producing a (nearly) input-CEP independent result --
 as long as the intensity and other parameters 
 are sufficiently well stabilized.

It is worth noting that 
 although these results show that CEP effects are negligible 
 in the regimes considered in this paper,
 we still see that the variances tend to increase for either 
 longer propagation distances
 or for higher pulse intensities.
Indeed, 
 for the high intensity case at $z=2.00$mm, 
 CEP fluctuations increase the log-variance to 
 about $-3$, 
 and have some non-negligible effect on
 the linear response to CEP fluctuations.
Clearly, 
 even a moderate further increase in intensity or propagation distance 
 would move the propagation into a regime where the
 train of broadened pulses is no longer coherent, 
 so that the Fang et al. scheme could no longer 
 generate phase stabilized pulse trains.

%
\section{Discussion}
\label{S-discussion}

This analysis has characterized the significance of 
 CEP sensitive processes in the pulse broadening process.
In particular, 
 $\sigma^2$ characterises the strength of these processes, 
 and highlights important features of extreme-SPM propagation.
However, 
 since the usual expectation is that the CEP sensitive THG terms
 have minimal effect  and the SPM term is dominant, 
 it is worth considering why the results here do not 
 always return an unambiguously linear phase response.

Firstly, 
 since the desired signal frequency will be in 
 the wings of the broadened spectrum, 
 small nonlinear effects can easily be significant.
Secondly, 
 the THG term applies to the entire spectrum, 
 including its low frequency wing.
Both these contributions are enhanced by 
 long propagation distances, 
 which provide more than enough opportunity for CEP sensitive
 effects to accumulate, 
 and then fold themselves back in to the propagating pulse.

One might wish to compare simulations 
 using a full $\chi^{(3)}E^3$ approach 
 against an SPM-only model.
Although non-trivial in standard PSSD, 
 it is possible to use wideband envelope techniques
 \cite{Kinsler-RN-2005pra,Genty-KKD-2007ox,Kolesik-M-2004pre,Kinsler-2007arXiv-envel}
 which allow the $\chi^{(3)}E^3$ term to be split efficiently
 into SPM and THG parts.
However, 
 this introduces a non-physical dispersion
 that averages over carrier cycles, 
 which removes the apparent value of such comparisons.
A more physical solution is to alter the dispersion 
 above (e.g.) $\omega \simeq 2\omega_0$, 
 to guarantee that there is no significant THG from the bulk of the
 pulse; 
 however this leaves in place CEP sensitive THG from the low-frequency
 wing of the pulse spectra.
Consquently it is not clear how to compare SPM and THG effects in a 
 way that makes physical sense.

%
\section{Conclusions}
\label{S-conclusions}

In this paper I have proposed measures to 
 assess the phase coherence of a pulse train 
 subject to propagation through a nonlinear dispersive medium.
I then used these to numerically investigate how pulse trains 
 broadened by extreme SPM could retain pulse-to-pulse 
 phase coherence, 
 as required by the Fang et al. \cite{Fang-K-2004ol} scheme.
The $\Delta_I$ measure allows us to set tolerances 
 in intensity fluctuations 
 that guarantee a level of output stability, 
 and the $\sigma^2_I$ measure tells us how reliable those
 tolerances will be.
The response to CEP ($\phi \rightarrow \Delta_\phi, \sigma^2_\phi$) 
 and pulse width ($\tau \rightarrow  \Delta_\tau, \sigma^2_\tau$) 
 fluctuations was also investigated.
Note that such trains of broadened pulses retain 
 remarkably good phase coherence in response to CEP fluctuations, 
 but that there still remains an underlying sensitivity to CEP changes.
Pulse with a wider bandwidth give a more robust spectral phase, 
 since extra bandwidth allows for rapid broadening to 
 the desired spectral range before 
 multi-wave cross-talk 
 can degrade (or destroys) the phase coherence.

Simulations of the type used in this paper along with the  
 $\sigma^2$ measures can provide useful insight into experimental 
 design.
However, 
 the simulation results presented here contain a great deal of fine detail, 
 so such simulations need to be carefully customized to the desired 
 experimental parameters.
This also suggests the potential for obtaining improved results 
 in experiment by making rather small (but very specific) 
 changes to the operating parameters.

Lastly, 
 the dominant process causing the loss of phase coherence in the 
 broadened pulse train is intensity fluctuations in the input; 
 varying the CEP of each pulse in the input train generally 
 has a much smaller effect.

%
\section*{Acknowledgments}

I acknowledge many useful discussions 
 with G.H.C. New, S.B.P. Radnor, and J.C.A. Tyrrell, 
as well as financial support from the EPSRC.

%

\begin{widetext}
\newpage
\end{widetext}

%
\section*{Appendix: CLEO'06 summary, 
 {\it ``Phase Retention in SPM Super-broadened Pulses''}}

\noindent
 P. Kinsler, S.B. Radnor, J.C.A. Tyrrell, G.H.C. New,

{\it
Pulses that have been super-broadened by a 
third-order nonlinearity
are frequently used when generating a phase stabilized output.
But how can such temporally mangled pulses 
retain any useful phase information?
}

\begin{figure}
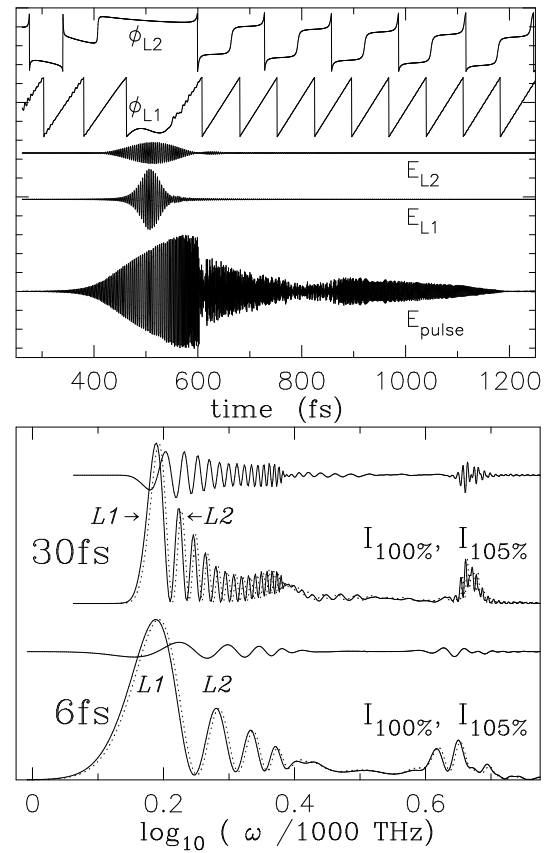
 
\begin{center}
 \includegraphics[angle=-90,width=0.80\columnwidth]{fig07a-30fs-time-1.00-combined.ps}\\
 \includegraphics[angle=-90,width=0.80\columnwidth]{fig07b-cfS.ps}
\end{center}
\caption{
Pulses at 786nm super-broadened in a sapphire-like medium to
 generate a lowest-frequency (1st) spectral lobe at $\sim 1200$nm
 to be used in the OPA stage of a phase stabilization 
 experiment \cite{Fang-K-2004ol,Baltuska-FK-2002prl}. 
\textbullet ~
TOP: 
Starting at the bottom, 
 this frame shows 
  the pulse's electric field ($E_{\textrm{\tiny pulse}}$), 
  a time-domain reconstruction of the electric field using 
  just only the frequency content of this 1st lobe (i.e. $E_{L1}$),
  and a similar reconstruction ($E_{L2}$) from just the second lobe.
The upper curves are the phase structure 
 $\phi_{L1}$ and $\phi_{L2}$ 
 first and second spectral lobes respectively.
\textbullet ~
BOTTOM:
Spectra of 30fs (top) and 6fs (bottom) pulses at 786nm after 
 similar amounts of spectral broadening, 
 which require different propagation lengths.
The nearly identical full and dotted lines compare the 
 spectra differing by the influence of a 5\%
 intensity variation. 
The other (offset) full lines shows the difference between those 
 very similar pairs of spectra at the two intensities.
}
\label{F-time-spect}
\end{figure}

We analyse recent experiments (such as 
\cite{Fang-K-2004ol,Baltuska-FK-2002prl}) in which
phase-stabilized idler pulses are generated in an optical parametric
amplifier (OPA) by combining a pump pulse with its super-broadened
replica.  The success of this self-stabilization experiment raises
interesting questions about how the severely distorted temporal
profile of the super-broadened signal can contain sufficient sensible
phase information to achieve the desired result.  
Our results demonstrate the mechanism by which phase order 
 survives amid the apparent temporal chaos of the pulse's
 electric field $E_{\textrm{\tiny pulse}}$.
Our analysis is informed by numerical simulations using 
 our recently-developed Pseudo-Spectral Space Domain (PSSD)
 technique \cite{Tyrrell-KN-2005jmo}, 
 and some results from one instructive example are shown in
 fig. \ref{F-time-spect}.

In traditional pulse propagation models based on 
 the complex envelope $A(t)$, 
 the nonlinear interaction term governing SPM appears as 
 $\chi^{(3)} |A|^2 A$.  
This is linear in the phase, 
 and might therefore be expected to be relatively well-behaved.  
However, one cannot model super-broadened pulses using envelopes, 
 even with the various corrections to envelope propagation
 that are available.  
One must therefore resort to a model capable of evolving 
 the carrier oscillations directly, 
 either the well known FDTD, or other methods like 
 PSSD\cite{Tyrrell-KN-2005jmo} or $G^\pm$ variables
 \cite{Kinsler-RN-2005pra}.
Given that the form of the nonlinear interaction is now 
 $\chi^{(3)} E^3$, 
 where $E(t)$ is the complete field, 
 it becomes unclear how the multiplicity of potential 
 three-wave interactions could generate and retain the 
 necessary phase information.  
The crucial questions we address are:

\noindent
\textbullet ~~
 How does phase information survive in 
 the presence of severe self-phase modulation and self-steepening?
\\
\textbullet ~~
Given that it does survive, how sensitive is the process to intensity
and phase fluctuations in the pump pulse? 
\\
\textbullet ~~
How does the number of
cycles in the pulse affect the behaviour?

Our simulations show firstly that a theoretical model of the
 SPM-OPA system described in \cite{Fang-K-2004ol,Baltuska-FK-2002prl} can 
 reproduce the character of the observed phase stability.  
Close analysis of the results shows that, 
 despite the mangled appearance of the super-broadened pulse,
 some parts of it retain an ordered (phase) structure, 
 and it is these sections that are used in the experiment
 (e.g. the 1st lobe in fig. \ref{F-time-spect}).

\begin{figure} 
\begin{center}
 \includegraphics[angle=-90,width=0.80\columnwidth]{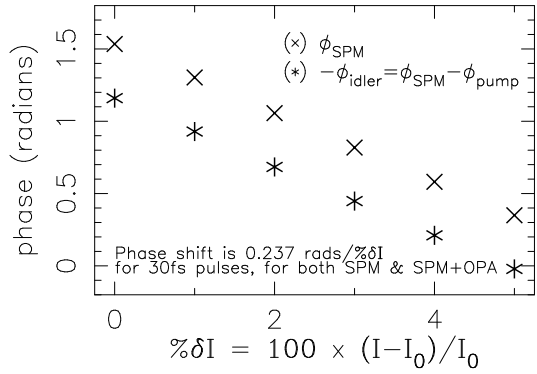}
\end{center}
\caption{
Phase stability as a function of percentage intensity variation 
 for an initial 30fs pulse, as per fig. \ref{F-time-spect}.
The $\phi_{SPM}$ data is extracted from the 1st spectral lobe
after the SPM broadening process.  
The $\phi_{idler}$ data uses the original pump pulse and the 
 SPM broadened copy in an 
 OPA stage to generate a phase-stabilized idler.
}
\label{F-phasestability}
\end{figure}

The investigation has also revealed more general rules 
 about the retention of phase information in strong SPM.  
The simulations have also enabled us to quantify  
 the sensitivity of the phase-stabilisation process 
 to fluctuations in the pump intensity, 
 a feature that bears directly on the viability of the technique 
 in the laboratory.  
The pump laser used in \cite{Fang-K-2004ol,Baltuska-FK-2002prl} 
 has an intensity stability
 to within 1\%, 
 leading to absolute phase fluctuations of $\sim 0.1\pi$ radians.
This compares remarkably well with our predicted $0.237$rad
 (see fig. \ref{F-phasestability}), 
 especially given that our model parameters do not closely match the  
 experiment.
Of course this match occurs because the phase stability 
 (w.r.t. intensity) is still linear, 
 and proportional to the broadening of the spectral component 
 phase-matched
 to the signal frequency selected by the set-up of the OPA stage.
Since {\em these} features are common to both 
 our model and the experiment, 
 the other differences of detail do little to 
 upset the comparison.

Lastly, we investigate the phase stability in the limit of 
 very few cycle pulses.
There are two effects here that can shift the phase variation 
 out of the linear regime seen for many cycle pulses.
The first is the nonlinear response time, 
 as modelled in \cite{Goorjian-C-2004ol}, 
 who used $\chi^{(3)}$ with a 1fs response time; 
 however, 
 note that  recent work \cite{Major-YNAS-2004ol}
 gives a faster response 
 (and thus will lead to smaller phase corrections).
The second is the strong non-SPM effects from the 
 $\chi^{(3)}$ interaction.
We show how these two nonlinear corrections to the phase properties
 affect the phase stability in the few-cycle limit, 
 and thus how they might affect experiments such as 
 \cite{Fang-K-2004ol,Baltuska-FK-2002prl} if pushed to those extreme limits.

\end{document}